\documentclass[reprint,superscriptaddress,amsmath, amssymb,aps,prd,notitlepage,longbibliography,floatfix,nofootinbib,twocolumn]{revtex4-1}

\usepackage{subfigure}
\usepackage{amsmath}
\usepackage{lipsum}
\allowdisplaybreaks[4]
\usepackage{cancel}
\usepackage{extarrows}
\usepackage{tensor}     
\usepackage{float}
\usepackage[final]{graphicx}   
\usepackage[
colorlinks=true,        
citecolor=blue,         
linkcolor=blue,         
urlcolor=blue           
]{hyperref}             
\usepackage{bm}         
\usepackage{xcolor}     
\usepackage{lipsum}
\usepackage{color}      
\usepackage[utf8]{inputenc} 
\usepackage[section]{placeins} 
\usepackage{appendix}
\usepackage{units}
\usepackage[capitalise]{cleveref}
\usepackage{units}
\usepackage{orcidlink}
\newcommand{\nc}{\newcommand*}

\nc{\xbar}{\bar{x}}
\nc{\rhoeq}{\rho_{\mathrm{eq}}}
\nc{\zeq}{z_{\mathrm{eq}}}
\nc{\tla}{\tilde{\lambda}}
\nc{\bt}{\beta}
\nc{\dt}{\delta}
\nc{\Dt}{\Delta}
\nc{\vj}{\vec{j}}
\nc{\vl}{\vec{l}}
\nc{\hx}{\hat{x}}
\nc{\hy}{\hat{y}}
\nc{\bj}{\bm{j}}
\nc{\mJ}{\mathcal{J}}
\nc{\mP}{\mathcal{P}}
\nc{\Msun}{M_\odot}
\nc{\app}{\approx}
\nc{\av}[1]{\langle #1 \rangle}
\nc{\eq}[1]{Eq.~\eqref{#1}}
\nc{\al}{\alpha}
\nc{\Xstar}{X_{\ast}}
\nc{\fpbh}{f_{\mathrm{pbh}}}
\nc{\vth}{\vec{\theta}}
\nc{\vla}{\vec{\lambda}}
\nc{\vd}{\vec{d}}
\nc{\Mmin}{M_{\mathrm{min}}}
\nc{\rmd}{\mathrm{d}}
\nc{\mmin}{{m_{\mathrm{min}}}}
\nc{\mmax}{{m_{\mathrm{max}}}}
\nc{\mR}{\mathcal{R}}
\nc{\tmR}{\tilde{\mathcal{R}}}
\nc{\s}{\sigma}
\nc{\ogw}{\Omega_{\mathrm{GW}}}
\nc{\addref}{[\textcolor{red}{add ref}] }
\nc{\Om}{\Omega}
\nc{\gm}{\gamma}
\nc{\Gm}{\Gamma}
\nc{\gpcyr}{\mathrm{Gpc}^{-3}\,\mathrm{yr}^{-1}}
\nc{\Eq}[1]{Eq.~\eqref{#1}}
\nc{\Fig}[1]{Fig.~\ref{#1}}
\nc{\Table}[1]{Table~\ref{#1}}
\nc{\lvc}{LIGO/Virgo} 
\nc{\Sec}[1]{Sec.~\ref{#1}}
\nc{\eg}{\textit{e.g.~}}
\nc{\SNR}{\mathrm{SNR}}
\nc{\be}{\mathbf{\epsilon}}
\nc{\bn}{\mathbf{n}}
\nc{\bd}{\mathbf{d}}
\nc{\ba}{\mathbf{a}}
\nc{\eps}{\epsilon}
\nc{\bnu}{\mathbf{\nu}}
\nc{\mb}{\mathbf}
\nc{\bbt}{\mathbf{t}}
\nc{\bth}{\mathbf{\theta}}
\nc{\bep}{\mathbf{\epsilon}}
\nc{\uni}{\mathrm{U}}
\nc{\logu}{\operatorname{\mathrm{log-U}}}
\nc{\RN}{\mathrm{RN}}
\nc{\BN}{\mathrm{BN}}
\nc{\GN}{\mathrm{GN}}
\nc{\mcN}{\mathcal{N}}
\nc{\GWB}{\mathrm{GW}}
\nc{\yr}{\mathrm{yr}}
\nc{\Am}{\mathcal{A}}
\nc{\Dm}{\mathcal{D}}
\nc{\Hm}{\mathcal{H}}
\nc{\sovast}{Soviet Ast.}

\nc{\mrm}{\mathrm}
\nc{\BE}{B\scriptsize{AYES}\normalsize{E}\scriptsize{PHEM}\normalsize  }

\nc{\Ostgw}{\Omega_{\mathrm{GW}}^{\mathrm{ST}}}
\nc{\Ottgw}{\Omega_{\mathrm{GW}}^{\mathrm{TT}}}
\nc{\Ovlgw}{\Omega_{\mathrm{GW}}^{\mathrm{VL}}}
\nc{\Oslgw}{\Omega_{\mathrm{GW}}^{\mathrm{SL}}}
\nc{\cosxi}{\beta}

\nc{\gmPL}{\gamma_{\mathrm{PL}}}
\nc{\APL}{A_{\mathrm{PL}}}

\def\({\left(}
\def\){\right)}
\def\[{\left[}
\def\]{\right]}

\def\e{\begin{equation}}
\def\q{\end{equation}}
\def\m{\begin{eqnarray}}
\def\n{\end{eqnarray}}
\nc{\red}[1]{\textcolor{red}{#1}}

\begin{document}

\title{The implications of inflation for the last ACT}

\author{Zhi-Chong Qiu\orcidlink{???}}
\email{qiuzhichong@itp.ac.cn}
\affiliation{Institute of Theoretical Physics, Chinese Academy of Sciences,Beijing 100190, China}
\affiliation{School of Physical Sciences, 
    University of Chinese Academy of Sciences, 
    No. 19A Yuquan Road, Beijing 100049, China}
\author{Ye-Huang Pang\orcidlink{0009-0004-5002-725X}}
\email{corresponding author: pangyehuang22@mails.ucas.ac.cn}
\affiliation{Institute of Theoretical Physics, Chinese Academy of Sciences,Beijing 100190, China}
\affiliation{School of Physical Sciences, 
    University of Chinese Academy of Sciences, 
    No. 19A Yuquan Road, Beijing 100049, China}
\affiliation{School of Fundamental Physics and Mathematical Sciences, Hangzhou Institute for Advanced Study, UCAS, Hangzhou 310024, China}
\author{Qing-Guo Huang\orcidlink{0000-0003-1584-345X}}
\email{corresponding author: huangqg@itp.ac.cn}
\affiliation{Institute of Theoretical Physics, Chinese Academy of Sciences,Beijing 100190, China}
\affiliation{School of Physical Sciences, 
    University of Chinese Academy of Sciences, 
    No. 19A Yuquan Road, Beijing 100049, China}
\affiliation{School of Fundamental Physics and Mathematical Sciences, Hangzhou Institute for Advanced Study, UCAS, Hangzhou 310024, China}


\begin{abstract}

We explored a parameterized slow-roll inflationary model within the $\Lambda$CDM framework, utilizing a combination of data from Planck 2018, ACT DR6, DESI DR2, and BICEP/Keck 2018 (P-ACT-LB-BK18). Additionally, we incorporated the SH0ES prior on $H_0$ (P-ACT-LB-BK18-$H_0$) to analyze the model within the early dark energy (EDE) framework. While the model with a potential $V(\phi)\propto \phi^\alpha$ for small values of $\alpha$ still fits the data, the Starobinsky $R^2$ inflation falls outside the $2\sigma$ region. On the other hand, in a self-consistent quantum theory of gravity, higher-order corrections to $R$ are typically anticipated. In response, we proposed a non-perturbative exponential $f(R)$ inflation model, wherein the subleading corrections beyond $R^2$ including terms like $R^3$ or $R^4$. Using numerical calculations and Markov Chain Monte Carlo (MCMC) analysis with the P-ACT-LB-BK18 data set, we demonstrate that this model can align well with the ACT-preferred value of the scalar spectral index. Additionally, within the early dark energy (EDE) framework, it accommodates greater deviations from the original Starobinsky $R^2$ inflation model when incorporating the SH0ES prior on $H_0$. 

\end{abstract}
\maketitle

\section{Introduction}

The standard inflationary paradigm \cite{Starobinsky:1980te,Guth:1980zm,Linde:1981mu,Albrecht:1982wi} elegantly solves the puzzles in the hot big bang model, such as the flatness problem, horizon problem and so on,  earning it widespread acceptance. More importantly, its prediction of a nearly scale-invariant spectrum of primordial scalar  perturbations is supported by observations of the cosmic microwave background (CMB), with the amplitude constrained to $\mathcal{P}_{\mathcal{R}} \simeq 2.1 \times 10^{-9}$~\cite{Planck:2018vyg}.

The Starobinsky $R^2$ inflation model provides predictions for the scalar spectral index $n_s$ consistent with the Planck 2018, which measures $n_s=0.9651\pm 0.0044$~\cite{Planck:2018vyg}. However, the latest data released from the Atacama Cosmology Telescope (ACT) have shifted the constraints on $n_s$ to a higher value \cite{ACT:2025fju,ACT:2025tim}. A joint analysis of ACT combined with Planck (P-ACT) yields $n_s = 0.9709\pm 0.0038$ \cite{ACT:2025fju,ACT:2025tim}, while further combining the Dark Energy Spectroscopic Instrument (DESI) measurements of baryon acoustic oscillations (BAO) \cite{DESI:2024mwx}, and the Bicep/Keck CMB experiments~\cite{BICEP:2021xfz} (P-ACT-LB$_{\text{DESI DR1}}$-BK18), places the Starobinsky $R^2$ inflation model at the $\sim 2\sigma$ boundary of constraints with a higher value of $n_s = 0.9743 \pm 0.0034$ \cite{ACT:2025fju,ACT:2025tim}, see~\cite{Kallosh:2025ijd} for a recent review.

The recent observational pressure on the Starobinsky $R^2$ inflation model has motivated theoretical efforts to refine inflationary models in light of the latest ACT observations~\cite{Kallosh:2025rni,Gialamas:2025kef,Gao:2025onc,Berera:2025vsu,Brahma:2025dio,Dioguardi:2025mpp,Aoki:2025wld,Dioguardi:2025vci,Salvio:2025izr,Rehman:2025fja,He:2025bli,Drees:2025ngb,Zharov:2025evb,Liu:2025qca,Yin:2025rrs,Gialamas:2025ofz,Yogesh:2025wak,Haque:2025uis,Yi:2025dms,Addazi:2025qra,Maity:2025czp,Peng:2025bws,Mondal:2025kur,Frolovsky:2025iao,Haque:2025uga,Pallis:2025nrv,Chakraborty:2025oyj,Kohri:2025lau,Heidarian:2025drk,Wolf:2025ecy,Okada:2025lpl,Gao:2025viy,Pal:2025ewf,Pallis:2025gii,Mohammadi:2025gbu,Zharov:2025zjg,Okada:2025nyd,Zhu:2025twm,Das:2025bws,Modak:2025bjv,Choi:2025qot,Aoki:2025ywt,Parvizi:2025sed,Yuennan:2025inm}, although some argue that this tension should be viewed with caution \cite{Ferreira:2025lrd,Linde:2025pvj}. On the other hand, from the perspective of an ultraviolet (UV) complete quantum theory of gravity, the higher curvature corrections to the Starobinsky $R^2$ inflation model are expected  \cite{Huang:2013hsb} and the Starobinsky $R^2$ inflation should lie in the swampland \cite{Lust:2023zql}.  Motivated by this theoretical consideration, we propose a non-perturbative exponential $f(R)$ inflation model that effectively incorporates subleading corrections beyond $R^2$, such as $R^3$ or $R^4$, in this article.

To obtain credible constraints on inflationary paradigms, the consistency of a cosmological model's predictions across diverse observational datasets is essential. The standard $\Lambda$CDM model faces several challenges, the most significant being the persistent Hubble tension \cite{Verde:2019ivm,DiValentino:2021izs,Abdalla:2022yfr} between the Planck 2018 ($H_0 = 67.27 \pm 0.60\;{\rm km\;s^{-1}\;Mpc^{-1}}$) \cite{Planck:2018vyg} and SH0ES ($H_0 = 73.17 \pm 0.86\;{\rm km\;s^{-1}\;Mpc^{-1}}$) \cite{2024ApJ...973...30B}. Furthermore, DESI BAO DR2, which favor a dynamical dark energy component evolving from phantom to quintessence~\cite{DESI:2025zgx}, come at the cost of intensifying the Hubble tension. However, it shows that, in  \cite{Pang:2025lvh}, the late-time $\Lambda$CDM model is still compatible with cosmological data including DESI BAO DR2 and the SH0ES prior on $H_0$ within the framework of the Early Dark Energy (EDE) model~\cite{Poulin:2018dzj,Poulin:2018cxd,Lin:2019qug,Niedermann:2019olb,Ye:2020btb,Karwal:2021vpk,Pang:2025lvh,Peng:2025tqt}. Therefore, we will constrain inflationary models using the P-ACT-LB-BK18 dataset within the $\Lambda$CDM framework, and the P-ACT-LB-BK18-$H_0$ dataset within the EDE framework, respectively.

The paper is organized as follows. In Sec. \ref{Sec_II}, we constrain a parameterized slow-roll inflation model within the $\Lambda$CDM framework using the P-ACT-LB-BK18, as well as within the EDE framework with P-ACT-LB-BK18-$H_0$, to access the observational preference for specific inflationary models. 
In Sec. \ref{Sec_III}, we propose a non-perturbative exponential $f(R)$ inflation model and place constraints on its parameters. Finally, in Sec. \ref{Sec_IV}, we summarize our findings and discuss their implications for inflationary models.

\section{Constraint on the parametrized inflation model for the last ACT}\label{Sec_II}

Following the parameterization scheme introduced in \cite{Huang:2007qz, Huang:2015cke}, the slow-roll parameter $\epsilon$ can be expressed in terms of the e-folding number before the end of inflation $N$ as
\begin{equation}
    \epsilon(N) = \dfrac{c/2}{(N + \Delta N)^p},
    \label{eps_para}
\end{equation}
where $c$ and $p$ are two parameters characterized inflation models. 
To ensure the condition $\epsilon(N=0) = 1$ is satisfied at the end of inflation, we introduce an offset $\Delta N$, given by
\begin{equation}
    \Delta N = \left(\dfrac{c}{2}\right)^{1/p}.
    \label{N_para}
\end{equation}
Using this parameterization, the scalar-to-tensor ratio $r$ and scalar spectral index $n_s$ for slow-roll inflation are given by
\begin{equation}
    r =
    \dfrac{8c}{(N + \Delta N)^p},
\end{equation}
\begin{equation}
    n_s 
    =1- \dfrac{c}{(N + \Delta N)^p} - \dfrac{p}{N + \Delta N}.
\label{r_ns}
\end{equation}
This parameterization effectively captures some key features of a broad class of inflationary models. For example,
\begin{itemize}
    \item Polynomial Inflationary Model: for the potential $V(\phi) \propto \phi^\alpha$, the parameters are given by $p = 1$ and $c = \alpha/2$.
    \item Starobinsky $R^2$ Inflation: 
    $p = 2$ and $c = 3/2$.
    \item Brane Inflation Model: for $d$-transverse-dimensional brane inflation model with potential $V(\phi) = V_0 ( 1-  (\mu_{
     \rm b}/\phi)^{d-2})$, the parameter $p$ is given by $p = 2(d-1)/d$. Taking D3-brane inflation as an example \cite{Kachru:2003sx}, the parameters take the form of $p = 5/3$ and $c \simeq 0$.
\end{itemize}
However, we note that this parameterization is not exhaustive of all inflation models.

We incorporate this parameterized inflation model into $\Lambda$CDM and EDE frameworks, performing a comprehensive Markov Chain Monte Carlo (MCMC) analysis to constrain the model parameters. For the EDE framework, we adopt an axion-like scalar potential
\e
V(\phi) = m^2 f^2 \[ 1-\cos \(\phi/f \) \]^{n_{\rm ede}},
\q
where $m$ is the scalar field mass, $f$ is the axion decay constant, and we fix $n_{\rm ede}=3$ \cite{Poulin:2018dzj, Poulin:2018cxd}. The potential allows the EDE component to behave like a cosmological constant at early times, and then rapidly dilute after recombination, as the field quickly rolls down its potential. This mechanism increases the Hubble rate around recombination, which reduces the sound horizon ($r_s$) at that epoch, thereby raising the inferred value of $H_0$ without violating the CMB constraints on angular sound horizon ($\theta_s$).

Within the $\Lambda$CDM framework, we constrain the model using the P-ACT-LB-BK18 dataset. For the EDE framework, we instead employ P-ACT-LB-BK18-$H_0$ dataset, which includes the SH0ES prior, $H_0 = 73.17 \pm 0.86\;{\rm km\;s^{-1}\;Mpc^{-1}} $.
We use \texttt{cobaya} \cite{Torrado:2020dgo} for MCMC sampling, interfaced with the \texttt{camb} code \cite{Lewis:1999bs,Fang:2008sn,Howlett_2012} for theoretical predictions under the $\Lambda$CDM model, while we use the \texttt{class$\_$ede}\footnote{ \url{https://github.com/mwt5345/class_ede}} code  \cite{Hill:2020osr} for EDE scenario.
We ensure reliable convergence for all chains by imposing a Gelman-Rubin criterion of $R - 1<0.05$ \cite{Gelman:1992zz}.

\begin{figure}[!htbp]
    \centering
    \includegraphics[width=0.9 \linewidth]{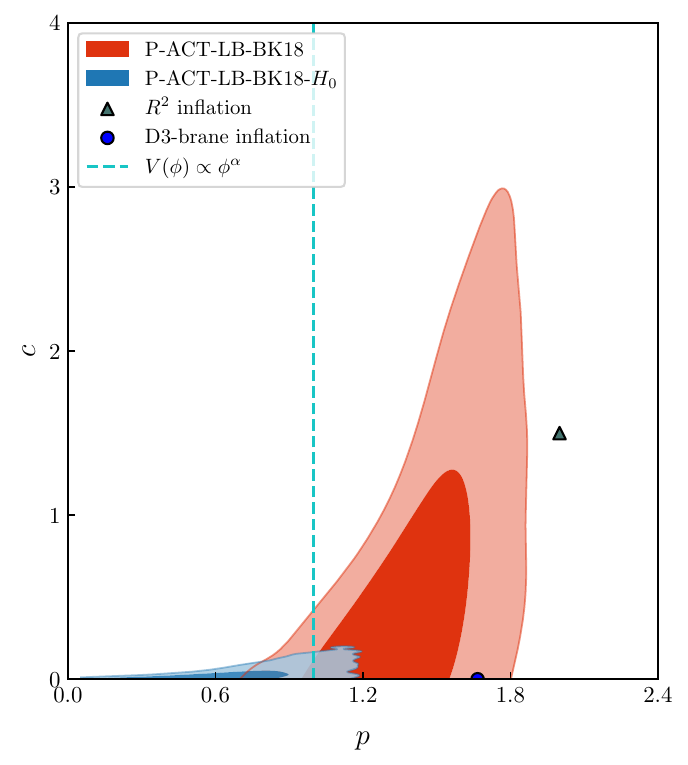}
    \caption{The constraints on $c$ and $p$ within $\Lambda$CDM and EDE frameworks by adopting  P-ACT-LB-BK18 and P-ACT-LB-BK18-$H_0$, respectively. Markers for $R^2$ inflation, D3-brane inflation and polynomial potential inflation are shown. }
    \label{fig:c_p}
\end{figure}

The constraints on $c$ and $p$ are presented in Fig.~\ref{fig:c_p}. 
Our analysis shows that P-ACT-LB-BK18 significantly tightens the bounds on $c$ and $p$ compared to previous results obtained within $\Lambda$CDM model~\cite{Huang:2015cke}. When incorporating EDE and using the P-ACT-LB-BK18-$H_0$ data set, the preferred value of $n_s$ increases, while the constraints on $c$ and $p$ become tighter.
Notably, the point corresponding to the $R^2$ inflation model lies outside the $2\sigma$ confidence region within both $\Lambda$CDM and EDE framework.
In contrast, the polynomial-potential inflation model still remains compatible with the data.

\section{Constraint on the non-perturbative exponential $f(R)$ inflation model for the last ACT}
\label{Sec_III}

Even though the plateau of the effective inflaton potential can drive the inflation in the Starobinsky $R^2$ inflation model, the curvature contributions of order higher than two are also expected in a consistent quantum theory of gravity \cite{Huang:2013hsb}. Additionally, the authors argued that the Starobinsky $R^2$ inflation lies in the swampland and concluded that it is therefore not a self-consistent model \cite{Lust:2023zql}.

The action for general $f(R)=R+F(R)$ gravity takes the form 
\e
S={M_p^2\over 2}\int d^4 x \sqrt{-g} f(R),
\label{ini_action}
\q
where $M_p$ denotes the reduced Planck scale. 
In \cite{Huang:2013hsb}, the function $f(R)$ is expressed in a polynomial form as  
\e
f(R)=R+{R^2\over 6M^2}+{\lambda_\alpha\over 2\alpha} {R^\alpha\over (3M^2)^{\alpha-1}}. 
\label{pfR}
\q
According to \cite{Huang:2013hsb}, a negative value of $\lambda_\alpha$ is necessary to achieve a higher scalar spectral index that aligns with ACT-preferred value. From a theoretical perspective, an $R^\alpha$ term with a negative coefficient is an unlikely result in a UV-complete theory, as $f(R)$ may turn negative for the highly curved geometry. Therefore, we propose a non-perturbative exponential $f(R)$ inflation model where $F(R)$ is given by 
\e
F(R) ={R^2\over 2\mu^2} \exp\[-\lambda (R/\mu^2)^n\],
\label{FR}
\q
with $\mu$ as an energy scale and $\lambda$ a dimensionless constant. This action reduces to the Starobinsky $R^2$-inflation in the limit of $\lambda\rightarrow 0$. By setting $\mu^2=3M^2$, the form of $F(R)$ in Eq.~(\ref{pfR}) can be taken as the leading and subleading term of expansion of $F(R)$ in Eq.~(\ref{FR}) for  $\lambda_\alpha=-\alpha \lambda$ with $n=\alpha-2$.

With auxiliary field $\chi=R$, one can rewrite the action (\ref{ini_action}) in an equivalent form
\e
S = \int d^4x \sqrt{-g}\[{{M_p^2}\over 2} \varphi R-U(\varphi) \]\,,
\label{varphi_action}
\q
where $\varphi=1+F_{,\chi}(\chi)$, hereafter the subscript denotes derivative. The potential is defined as
\e
U(\varphi) = {M_p^2 \over 2}\[(\varphi-1)\chi(\varphi)-F(\chi(\varphi)) \],
\q
which requires $F_{,\chi\chi} \ne 0$. 
By using a conformal transformation $g_{\mu\nu}^E = \varphi g_{\mu\nu}$, and introducing the canonical scalar field by
\e
\phi = \sqrt{3\over 2}M_p \ln\varphi,
 \label{phi_and_varphi}
\q
the action can be rewritten in the Einstein frame by
\e
S_E = \int d^4 x \sqrt{-g_E}\[{M_p^2 \over 2}R_E - {1\over 2} g_E^{\mu\nu} \partial_\mu \phi \partial_\nu \phi - V(\phi)  \],
\label{inflation_action}
\q
where $V(\phi) = U(\varphi)/\varphi^2$. For simplicity, we shall omit the subscript “E” in what follows and work entirely in the Einstein frame.

The slow-roll parameters $\epsilon$ and $\eta$ can be expressed in terms of $\varphi$ as
\e
\epsilon = {M_p^2\over 2}\left({V_{,\phi}\over V} \right)^2 = {\varphi^2\over 3}\({V_{,\varphi}\over V} \)^2,
\q
and 
\e
\eta = M_p^2{V_{,\phi\phi}\over V} = {2\over 3} {\varphi V_{,\varphi} + \varphi^2 V_{,\varphi\varphi} \over V }.
\q
The e-folding number before the end of inflation $N$ is related to the initial field value $\varphi_N$ by
\e
N\simeq {1\over M_p}\int_{\phi_{\text{end}}}^{\phi_N} {V\over V_{,\phi}}d\phi = {3\over 2}\int_{\varphi_{\text{end}}}^{\varphi_N} { V\over V_{,\varphi} \varphi^2} d\varphi,
\q
where $\varphi_{\text{end}}$ is determined by the condition $\epsilon(\varphi_{\text{end}})\simeq1$. The primordial scalar spectrum is given by 
\e
\mathcal{P}_{\mathcal{R}} = {V\over 24\pi^2  M_p^4 \epsilon}\bigg|_{\varphi_N} \simeq {\mu^2 N^2\over 72 \pi^2 M_p^2}.
\q
Based on the Planck normalization $\mathcal{P}_{\mathcal{R}}\simeq 2.1\times 10^{-9} $ \cite{Planck:2018vyg}, we can determine the energy scale $\mu$ for a given $N$.

\begin{figure}[htbp]
  \centering
  \includegraphics[width=0.4\textwidth]{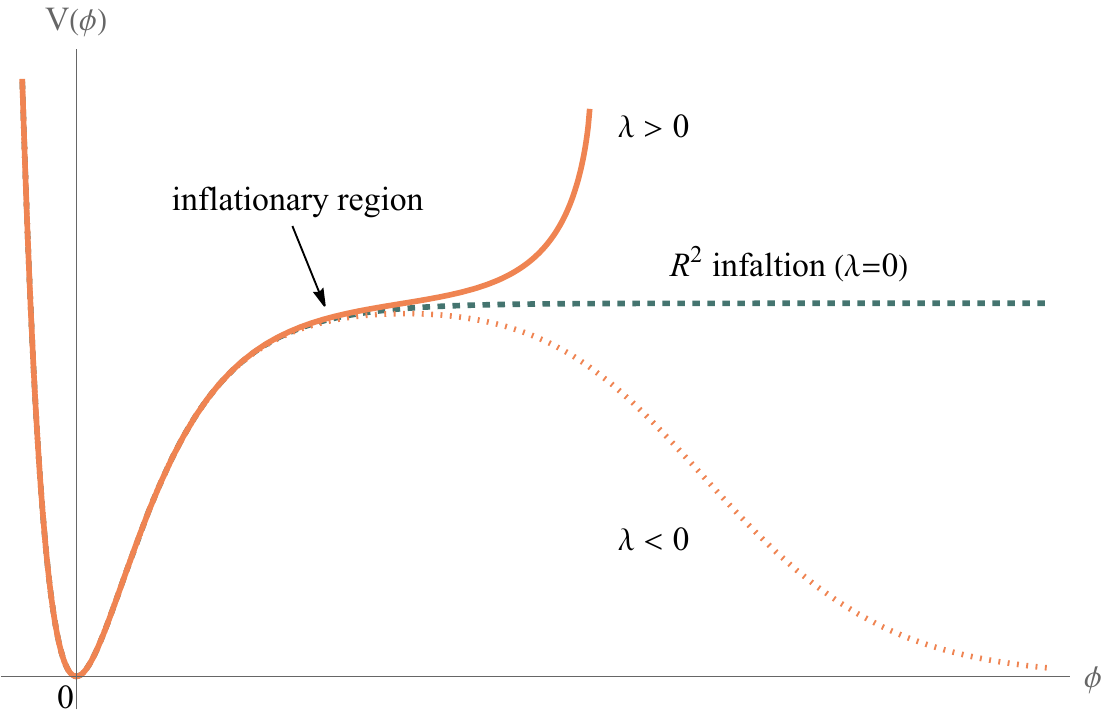}
  \caption{The potential of $\phi$ .}
  \label{Vphi_potential}
 \end{figure}

A general analytic expression for $\chi(\varphi)$ is unavailable when $n$ is a positive integer and $\lambda\ne0$ for Eq. (\ref{FR}). Therefore, we evaluate the potential $V(\phi)$ numerically, as shown in Fig.~\ref{Vphi_potential}, and the resulting potential shape is similar to the polynomial $f(R)$ inflation model (\ref{pfR}) shown in \cite{Huang:2013hsb}.

For a qualitative understanding of the parameter $\lambda$, the second derivative of $F(R)$ is given by
 \e
 F_{,RR}(R) = {\exp\[-x/n\]\over 2\mu^2}\[x^2-\(n+3\)x+2  \],
 \q
where $x= n\lambda (R/\mu^2)^n$, and
\e
\varphi  =1+ {\exp\[-x/n\] R \over 2\mu^2} \( 2-x \).
\label{varphi-1}
\q

For $\lambda>0$, the first critical point $F_{,RR}(R)=0$  occurs at\footnote{For $\lambda<0$, $F_{,RR}$ remains positive since the condition $x<0$ always holds.}
\e
x_c = {n+3- \sqrt{n^2+6n+1} \over 2},
\q
corresponding to the Ricci scalar $R_c = \mu^2\({x_c\over n\lambda}\)^{1/ n} $. Once we substitute $R_c$ into Eq. (\ref{varphi-1}), $\lambda$ cannot be too large, as a sufficiently long plateau-like region in the potential is required to generate a large enough $N$. In addition, $F_{,R}(R)$ exhibits a maximum at $R_c$ where $F_{,RR}(R)$ changes sign, implying an upper bound on the scalar field $\varphi\le 1+F_{,R}(R_c)$, thereby avoiding tachyonic instability $f_{,RR}(R)<0$. The ghost-free condition $f_{,R}(R)>0$ is automatically satisfied, since $f_{,R}(R)=0$ corresponding to $\phi \to -\infty$ is dynamically inaccessible.  

For $\lambda<0$, the calculations of perturbative polynomial model (\ref{pfR}) indicate that the predicted scalar spectral index $n_s$ is lower than that of the  Starobinsky $R^2$ inflation \cite{Huang:2013hsb}, and would therefore be largely inconsistent with the latest ACT data.

\begin{figure}[htbp]
  \centering
  \includegraphics[width=0.45\textwidth]{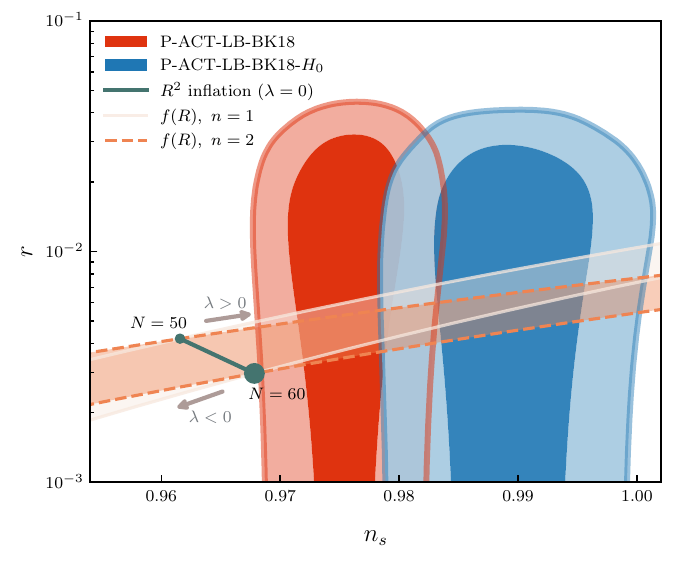}
  \caption{The predictions of the non-perturbative exponential $f(R)$ inflation model versus the constraints from P-ACT-LB-BK18 and P-ACT-LB-BK18-$H_0$ within the $\Lambda$CDM and EDE frameworks, respectively. Notably, the shaded region on the right of $R^2$ inflation line corresponds to $\lambda>0$, while the shaded region on the left corresponds to $\lambda<0$, which is strongly disfavored.}
  \label{fR_r_ns}
 \end{figure}

We numerically compute $n_s$ and $r$ for $n=1$ and $n=2$ with different choices of $N$ and $\lambda$ for the model (\ref{FR}), as shown in Fig.~\ref{fR_r_ns}. The contours represent the constraints obtained within the same cosmological background framework introduced in \Sec{Sec_II}, i.e. $\Lambda$CDM or EDE. When varying the e-folding number before the end of inflation within $50 \le N \le 60$, the results show that only for $N\simeq 60$ the Starobinsky $R^2$ inflation model ($\lambda=0$) lies near the edge of $2\sigma$ confidence region within $\Lambda$CDM framework, while the model falls outside the $2\sigma$ region within EDE framework, consistent with the results in \Sec{Sec_II}. Furthermore,  Fig.~\ref{fR_r_ns} shows that the predictions of the non-perturbative exponential $f(R)$ inflation model are consistent with the data for a range of $\lambda$ values.  

\begin{figure}[!htbp]
  \centering
  \begin{subfigure}
    \centering
    \includegraphics[width=0.35\textwidth]{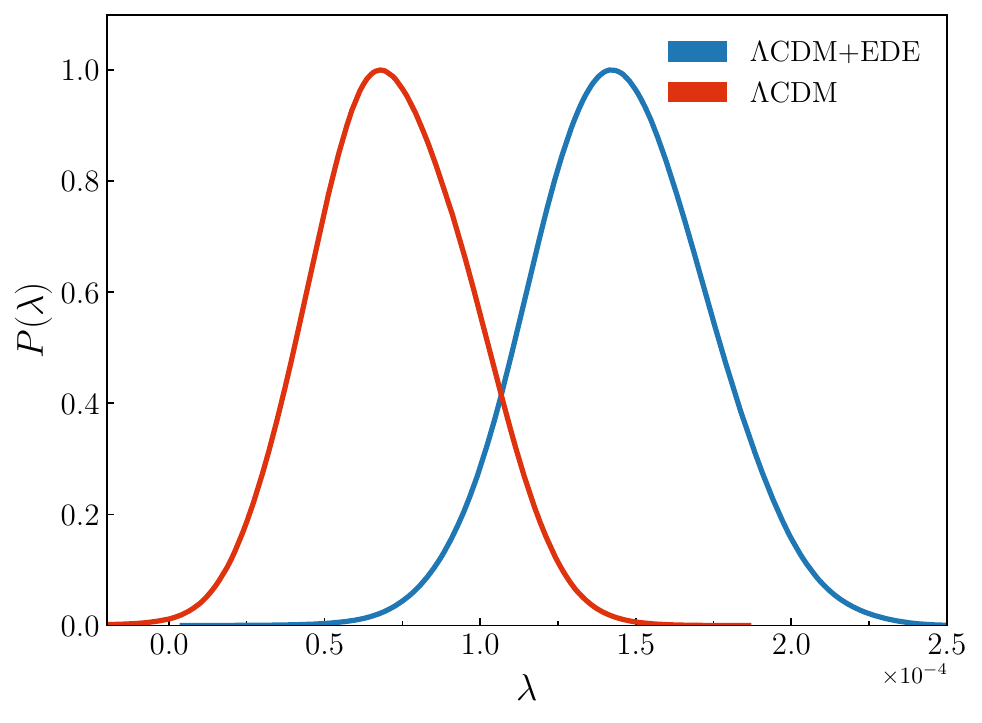}
  \end{subfigure}
  \hfill
  \begin{subfigure}
    \centering
    \includegraphics[width=0.35\textwidth]{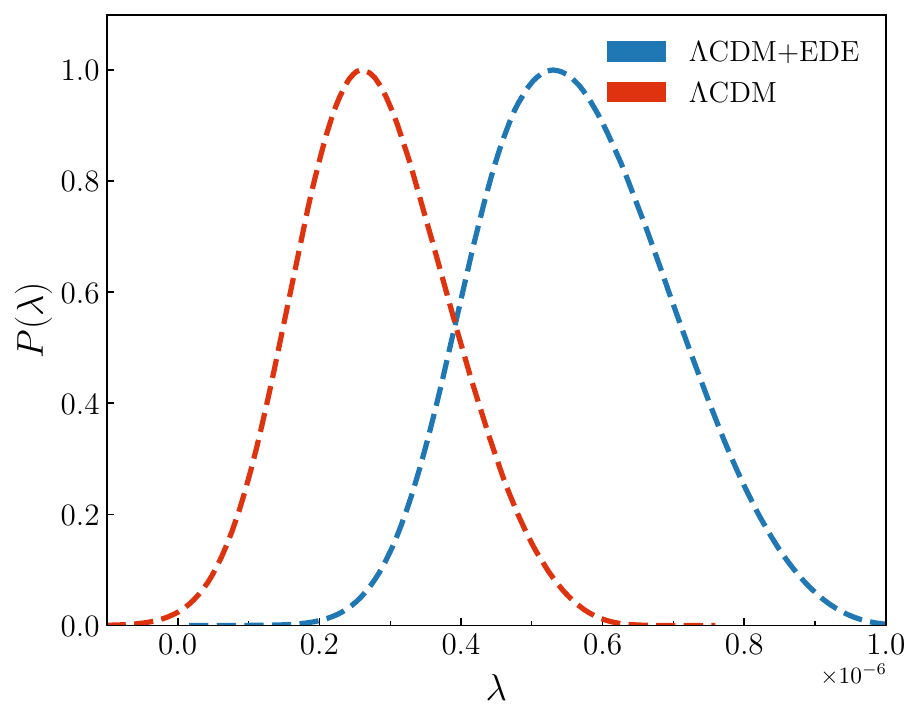}
  \end{subfigure}
\caption{Constraints on $\lambda$ for $n=1$ (solid line) and $n=2$ (dashed line) in the  non-perturbative exponential $f(R)$ gravity inflation model by using  P-ACT-LB-BK18 and P-ACT-LB-BK18-$H_0$ within the $\Lambda$CDM and EDE frameworks respectively.}
  \label{fig_results}
\end{figure}

We obtain quantitative constraints on $\lambda$ using MCMC analyses within both cosmological frameworks, which yield the results shown in Fig.~\ref{fig_results}. In both frameworks, the posterior distributions favor positive $\lambda$, with larger values preferred in the EDE scenario due to its tendency toward a higher $n_s$; whereas, the Starobinsky $R^2$ inflation model and negative $\lambda$ values are statistically disfavored.

\section{Conclusion and discussion} 
\label{Sec_IV}
The latest cosmological data from ACT DR6 introduce a notable tension with the Starobinsky $R^2$ inflation model. In this work, we assessed the implications of the P-ACT-LB-BK18 within $\Lambda$CDM and P-ACT-LB-BK18-$H_0$ within EDE framework constraining inflationary scenarios. Using two parameters $c,\ p$ description for the slow-roll parameter $\epsilon$, we examined typical inflationary models such as polynomial inflationary model,  Starobinsky $R^2$ inflation and so on. Our analysis indicates that these new datasets significantly tighten the constraints compared to previous results~\cite{Huang:2015cke}. Notably, the Starobinsky $R^2$ inflation model lies outside the $2\sigma$ confidence region for $\Lambda$CDM, whereas models with potentials $V(\phi)\propto \phi^\alpha$ remain compatible with the data for small values of $\alpha$.

On the other hand, higher-order curvature terms are generally anticipated in a consistent quantum theory of gravity and may address the existing tension between the Starobinsky $R^2$ inflation model and latest ACT data. Motivated by this theoretical consideration, we propose a non-perturbative exponential $f(R)$ inflation model, focusing on the typical cases $n=1$ and $n=2$ corresponding to $R^3$ and $R^4$ corrections in \cite{Huang:2013hsb}. Through numerical calculations and MCMC analyses, our results show that this model can shift the prediction of $n_s$ to higher values compared to Starobinsky $R^2$ inflation in better agreement with the ACT-preferred region within both the $\Lambda$CDM and EDE frameworks, with the latter favoring larger $n_s$ and a correspondingly shifted range of $\lambda$. 

Looking ahead, the upcoming data from CMB-S4 \cite{CMB-S4:2016ple}, Simons Observatory \cite{SimonsObservatory:2018koc}, and LiteBIRD \cite{LiteBIRD:2022cnt} are expected to provide tighter constraints on $r$ and $n_s$, potentially enabling more stringent tests of inflationary models. Our results indicate that non-perturbative exponential $f(R)$ models offer a viable mechanism to reconcile higher-order curvature corrections with current CMB observations, suggesting that such non-perturbative extensions of Starobinsky $R^2$ inflation could be further tested and constrained with these forthcoming observations.

\textit{Acknowledgements.}
QGH is supported by the grants from NSFC (Grant No.~12475065, 12447101) and the China Manned Space Program with grant no. CMS-CSST-2025-A01.

\bibliography{refs}
\end{document}